\documentclass{npcs}
\usepackage{graphics} 
\usepackage{epsfig}
\usepackage{amsmath, amssymb}
\usepackage{axodraw}
\usepackage{multicol}
\begin{document}

\topmargin = -30mm

\runauthor{T. V. Shishkina, V. V. Makarenko}
\runtitle{
The investigation of spin effects in photon production with fermion pair
in $\gamma \gamma$-collisions
}

\begin{topmatter}
\vspace{5cm}
\title{The investigation of spin effects in photon production with fermion pair
in $\gamma \gamma$-collisions}
\author{Author T.V. Shishkina}
\institution{NC PHEP BSU}
\address{153 Bogdanovitcha str.,220040 Minsk, Belarus}
\email{shishkina@hep.by}
\author{Author V.V. Makarenko}
\institution{NC PHEP BSU}
\address{153 Bogdanovitcha str.,220040 Minsk, Belarus}
\email{makarenko@hep.by}
\vspace{2cm}
\begin{abstract}
The exclusive reaction $ \gamma\gamma\to f \bar{f} \gamma$
is considered as the
possible calibration process on the linear photon collider.
We analyse the energy spectrum and total cross section dependence
on the detector parameters at various initial beam helicities.
\end{abstract}	
\end{topmatter}



\mathchardef\vm="117
\mathchardef\um="11D
\mathchardef\E="245
\mathchardef\Mom="250
\def\z1{z{}'}
\def\t1{t{}'}
\def\u1{u{}'}
\def\s1{s{}'}
\def\m2{m^2}
\def\d12{\frac{1}{2}}
\def\lfr#1#2{\ln{\frac{#1}{#2}}}

\def\ub{\bar{\um}}
\def\Jint#1{\mathcal{J} ( {#1} )}
\def\Iint#1#2{\int\limits_{0}^{\ub}{\Jint{#1}} {#2} {d \um}}
\def\Iintl#1#2{\int\limits_{2 m \lambda}^{\ub}{\Jint{#1}} {#2} {d \um}}

\def\bracket#1{\left({#1}\right)}
\def\bra#1{\bracket{#1}}
\def\spr#1#2{\, #1 \! \cdot \! #2 \,}
\def\ddx#1{\frac{1}{#1}}

\newpage

\section{Introduction}

Linear lepton colliders of the nearest future provide
the possibility of investigation of
photons collisions at energies and luminosities
the same to
those in $e^+e^-$ collisions \cite{gg_proposal}.
The photon linear colliders have the great physical potential
(Higgs ans SUSY particles searching,
study of anomalous gauge boson couplings
and hadronic structure of photons etc., \cite{tdr}).
And performing of this set of investigations
requires
the of photon beam calibration as well.
For this purpose one traditionally use some of the well-known
and precisely calculated reactions
(see, for example, $\gamma\gamma\to 2 f, 4 f$, \cite{gg_2f, gg_4f}).
The difficulties appear in the calibration of
photon beams of similar helicity
due to the small magnitude of cross sections
of the most QED processes.

For this purpose we consider the exclusive reaction $\gamma\gamma\to
f\bar{f}\gamma$ as the possible candidate 
for a calibration
channel on a linear photon collider.

The two various helicity
configuration of the $\gamma \gamma$-system leads to the different spectra
of final particles
and requires the two mechanisms of beam calibration.
We have analysed the behaviour of the
$\gamma\gamma\to f\bar{f}\gamma$ reaction
on beams with different helicities
as a function of the parameters of detectors.
We have performed the detail comparison of it's cross sections on $\gamma^+\gamma^+$- 
(total helicity $J=0$) and $\gamma^+\gamma^-$-beams  ($J=2$).
The question to be answered is does the process
on "$++$" beams can have the same cross section as one on "$+-$"
beams. We have outlined the conditions that greatly restrict the
observation of the process on "$J=2$" beams, remaining the "$J=0$" cross
section almost unchanged.


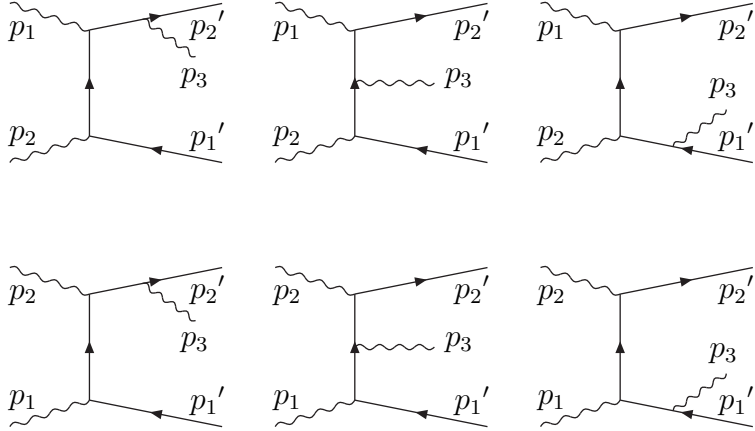
\begin{figure}[h]
\begin{picture}(300,200)(-90,0)
\Photon(10, 20)(40, 30){1}{4}
\Photon(10, 80)(40, 70){1}{4}
\ArrowLine(90, 20)(40, 30)
\ArrowLine(40, 30)(40, 70)
\ArrowLine(40, 70)(90, 80)
\Photon(60, 74)(80, 60){1}{4}
\Text(15,30)[c]{\normalsize $p_1$}
\Text(15,70)[c]{\normalsize $p_2$}
\Text(80,52)[c]{\normalsize $p_3$}
\Text(85,30)[c]{\normalsize $p_1{}'$}
\Text(85,72)[c]{\normalsize $p_2{}'$}

\Photon(110, 20)(140, 30){1}{4}
\Photon(110, 80)(140, 70){1}{4}
\ArrowLine(190, 20)(140, 30)
\ArrowLine(140, 30)(140, 70)
\ArrowLine(140, 70)(190, 80)
\Photon(140, 50)(170, 50){1}{4}
\Text(115,30)[c]{\normalsize $p_1$}
\Text(115,70)[c]{\normalsize $p_2$}
\Text(180,52)[c]{\normalsize $p_3$}
\Text(185,30)[c]{\normalsize $p_1{}'$}
\Text(185,72)[c]{\normalsize $p_2{}'$}

\Photon(210, 20)(240, 30){1}{4}
\Photon(210, 80)(240, 70){1}{4}
\ArrowLine(290, 20)(240, 30)
\ArrowLine(240, 30)(240, 70)
\ArrowLine(240, 70)(290, 80)
\Photon(260, 26)(280, 40){1}{4}
\Text(215,30)[c]{\normalsize $p_1$}
\Text(215,70)[c]{\normalsize $p_2$}
\Text(280,48)[c]{\normalsize $p_3$}
\Text(285,30)[c]{\normalsize $p_1{}'$}
\Text(285,72)[c]{\normalsize $p_2{}'$}

\SetOffset(0,100)
\Photon(10, 20)(40, 30){1}{4}
\Photon(10, 80)(40, 70){1}{4}
\ArrowLine(90, 20)(40, 30)
\ArrowLine(40, 30)(40, 70)
\ArrowLine(40, 70)(90, 80)
\Photon(60, 74)(80, 60){1}{4}
\Text(15,30)[c]{\normalsize $p_2$}
\Text(15,70)[c]{\normalsize $p_1$}
\Text(80,52)[c]{\normalsize $p_3$}
\Text(85,30)[c]{\normalsize $p_1{}'$}
\Text(85,72)[c]{\normalsize $p_2{}'$}

\Photon(110, 20)(140, 30){1}{4}
\Photon(110, 80)(140, 70){1}{4}
\ArrowLine(190, 20)(140, 30)
\ArrowLine(140, 30)(140, 70)
\ArrowLine(140, 70)(190, 80)
\Photon(140, 50)(170, 50){1}{4}
\Text(115,30)[c]{\normalsize $p_2$}
\Text(115,70)[c]{\normalsize $p_1$}
\Text(180,52)[c]{\normalsize $p_3$}
\Text(185,30)[c]{\normalsize $p_1{}'$}
\Text(185,72)[c]{\normalsize $p_2{}'$}

\Photon(210, 20)(240, 30){1}{4}
\Photon(210, 80)(240, 70){1}{4}
\ArrowLine(290, 20)(240, 30)
\ArrowLine(240, 30)(240, 70)
\ArrowLine(240, 70)(290, 80)
\Photon(260, 26)(280, 40){1}{4}
\Text(215,30)[c]{\normalsize $p_2$}
\Text(215,70)[c]{\normalsize $p_1$}
\Text(280,48)[c]{\normalsize $p_3$}
\Text(285,30)[c]{\normalsize $p_1{}'$}
\Text(285,72)[c]{\normalsize $p_2{}'$}

\end{picture}

\caption{
Diagrams for the process $ \gamma\gamma\to f \bar{f}+\gamma$.
}
\end{figure}

We consider the process
\begin{eqnarray}
\gamma(p_1, \lambda_1)+\gamma(p_2, \lambda_2) \to f(p_1{}', e_1{}') + \bar{f}(p_2{}', e_2{}') + \gamma(p_3, \lambda_3),
\end{eqnarray}
where $\lambda_{i}$ and $e_{i}{}'$ are photon and fermion helicities.

We use the following set of invariants:
\begin{eqnarray}\nonumber
\begin{array}{lll}
s = {\left(p_1+p_2\right)}^2 = 2 \spr{p_1}{p_2},
&
t = {\left(p_2{}'-p_2\right)}^2,
&
u = {\left(p_2{}'-p_1\right)}^2,
\\
s{}' = {\left(p_1{}'+p_2{}'\right)}^2,
&
t{}' = {\left(p_1{}'-p_1\right)}^2,
&
u{}' = {\left(p_1{}'-p_2\right)}^2,
\\
\um = 2 \spr{p_1{}'}{p_3},
& 
\vm = 2 \spr{p_2{}'}{p_3},
\\
z = 2 \spr{p_1}{p_3},
& 
z{}' = 2 \spr{p_2}{p_3}.
\end{array}
\end{eqnarray}

We denote the final-state photon energy by $w$,
the polar angle
(between any final and initial particles in c.m.s.)
by $\Theta$
and the angle between any of final particles by $\varphi$.
For the differential cross-section we introduce the normalized
final-state photon energy (c.m.s. is used) $
x=w\slash\sqrt{s}$.
The differential cross section ${d\sigma}\slash{d x}$
appears to be the energy spectrum of
final-state photons.
The total helicity of the $\gamma \gamma$ system is denoted by $J$.

\section{Calculation}

We consider the cross section
%
%
\begin{eqnarray}\nonumber
\sigma =\int \ddx{2 s} {\left|M (\lambda_1,\lambda_2,e_1{}',e_2{}',\lambda_3)\right|}^{2} d \Gamma,
\end{eqnarray}
where the phase-space volume element is defined by
\begin{eqnarray}\nonumber
d \Gamma =
\frac{d^{3} p_1'}{\bra{2 \pi}^{3} 2 \E_1'}
\cdot \frac{d^{3} p_2'}{\bra{2 \pi}^{3} 2 \E_2'}
\cdot \frac{d^{3} p_3}{\bra{2 \pi}^{3} 2 p_3^0}
\cdot {\bra{2 \pi}^{4}} \delta \bra{p_1+p_2-p_1'-p_2'-p_3}.
\end{eqnarray}

The process $ \gamma\gamma\to f \bar{f}\gamma$ is the pure QED
reaction in the Born approximation.
Using the method of helicity amplitudes \cite{ha}, we calculate
matrix elements:
\begin{eqnarray}\label{br3}
\left|M^{+--++}\right| = 2 e^3 \spr{p_2'}{p_2}
\sqrt{\frac{\spr{p_1'}{p_2'}}{\spr{p_1'}{p_3} \spr{p_2'}{p_3} \spr{p_1'}{p_1} \spr{p_2'}{p_1}}}
= e^3
\sqrt{\frac{8 s t^{2}}{\vm \um t{}' u}}.
\end{eqnarray}

All the other non-vanishing amplitudes are
obtained from $\left|M^{+--++}\right|$
by using C, P, Bose and crossing (between final and initial particles) symmetries:
$$
\qquad\qquad
d \sigma^{+-+--} =
d \sigma^{+--++} {}_{\mid_{1\leftrightarrow 2}}, \;(P+Bose)
$$
$$
\quad
d \sigma^{+-+-+} =
d \sigma^{+--++} {}_{\mid_{1'\leftrightarrow 2'}}, \;(C)
$$ $$
\qquad\qquad\quad
d \sigma^{+--+-} =
d \sigma^{+--++} {}_{\mid_{1\leftrightarrow 2 \atop
1'\leftrightarrow 2'}}, \;(CP+Bose)
$$
$$
\qquad\qquad\qquad
d \sigma^{+++--} =
d \sigma^{+--++} {}_{\mid_{3\leftrightarrow 2 \atop
1'\leftrightarrow 2'}}, \;(C+crossing)
$$
$$ 
\quad
d \sigma^{++-+-} =
d \sigma^{+++--} {}_{\mid_{1'\leftrightarrow 2'}},
\;(C)
$$
$$ 
\quad
d \sigma^{-\lambda_1,-\lambda_2,-e_1{}',-e_2{}',-\lambda_3} =
d \sigma^{\lambda_1,\lambda_2,e_1{}',e_2{}',\lambda_3}.
\;(P)
$$

Due to no final-state polarizations can be measured we summarize
over all the final particles helicities.

The further integration is performed numerically using the
Monte-Carlo method \cite{mc}.

\section{Results}

One of the main purposes of the linear photon collider
is the $s$-channel of the Higgs boson production
at energies about $\sqrt{s}=120 GeV$ \cite{gg_h}.
Hence we use this value of c.m.s. energy in our analysis.

We have performed calculations
for various experimental restrictions on
the parameters of final particles.
The events are not detected if
$\Theta$, $\varphi$ or $w$ are below the threshold values.
The default minimal values are $\Theta_{min} = 7^{o}$,
$\varphi_{min} = 3^{o}$,
$w_{min} = 1 GeV$,
the minimal fermion energy was also assumed to be $1 GeV$.

\begin{figure}[h!]
\leavevmode
\begin{minipage}[b]{.475\linewidth}
\centering
\includegraphics[width=\linewidth, height=3.8in, angle=0]{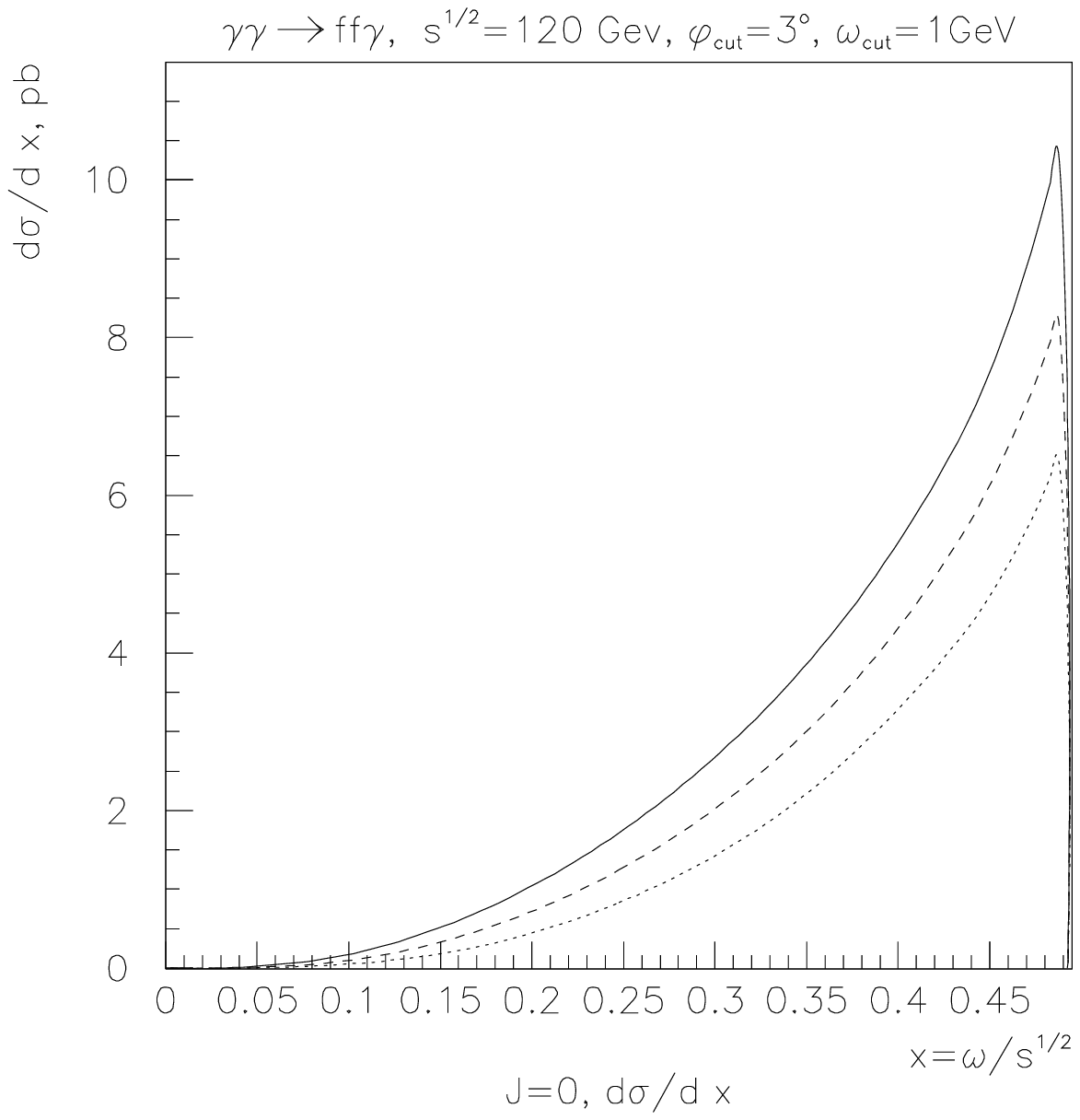}
\label{d_1_120}
\end{minipage}\hfill
\begin{minipage}[b]{.475\linewidth}
\centering
\includegraphics[width=\linewidth, height=3.8in, angle=0]{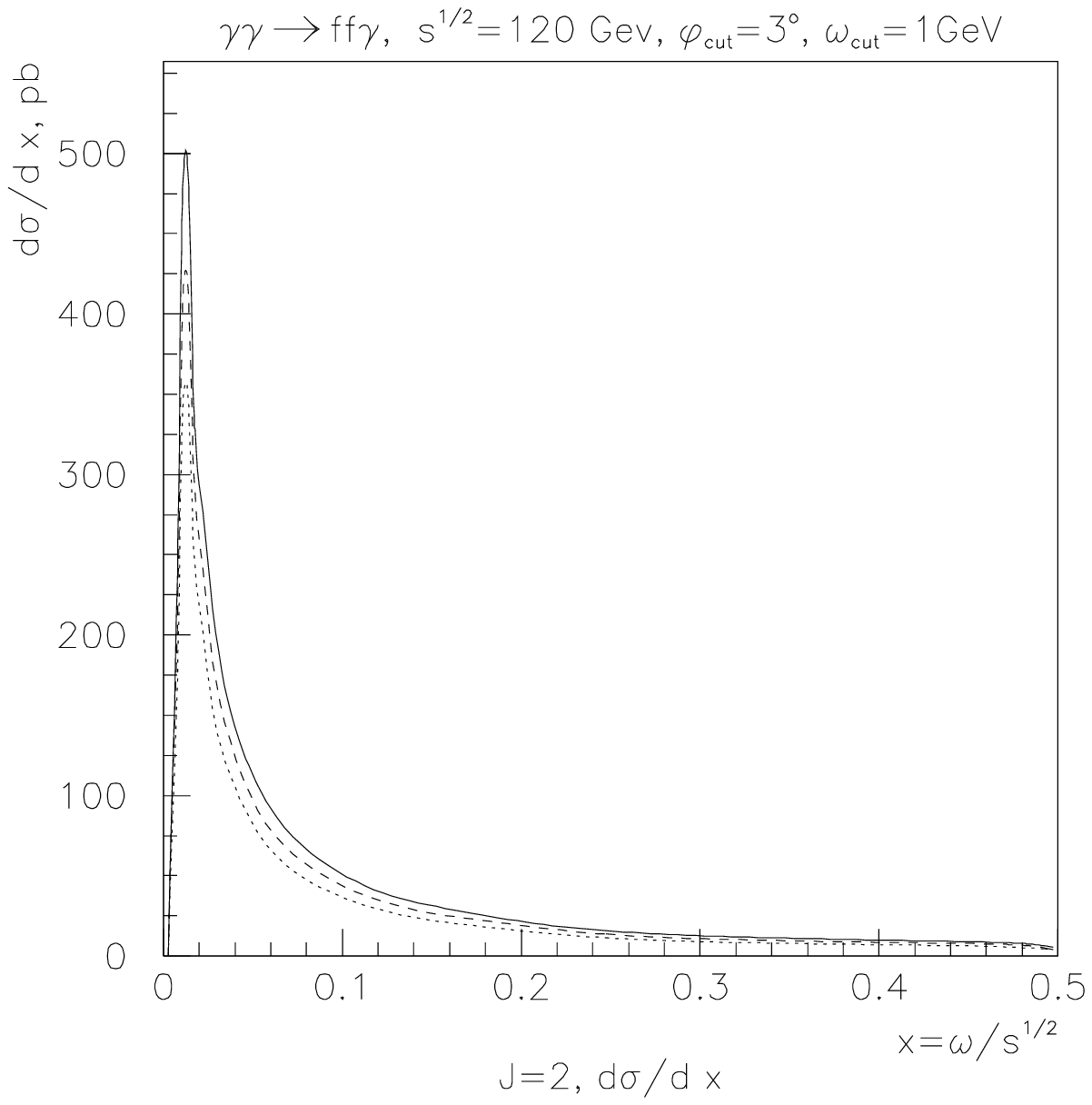}
\label{d_2_120}
\end{minipage}
\caption{
The cross section ${d\sigma}\slash{d x}$ for $\sqrt{s} = 120 GeV$.
The cuts are: $5^o,7^o$ and $10^o$ on the polar angle,
$3^o$ on the collinear angle and $1 GeV$ on
the minimal final-state particle energy.
}\label{f1}
\end{figure}

\begin{figure}[h!]
\leavevmode
\begin{minipage}[b]{.475\linewidth}
\centering
\includegraphics[width=\linewidth, height=3.8in, angle=0]{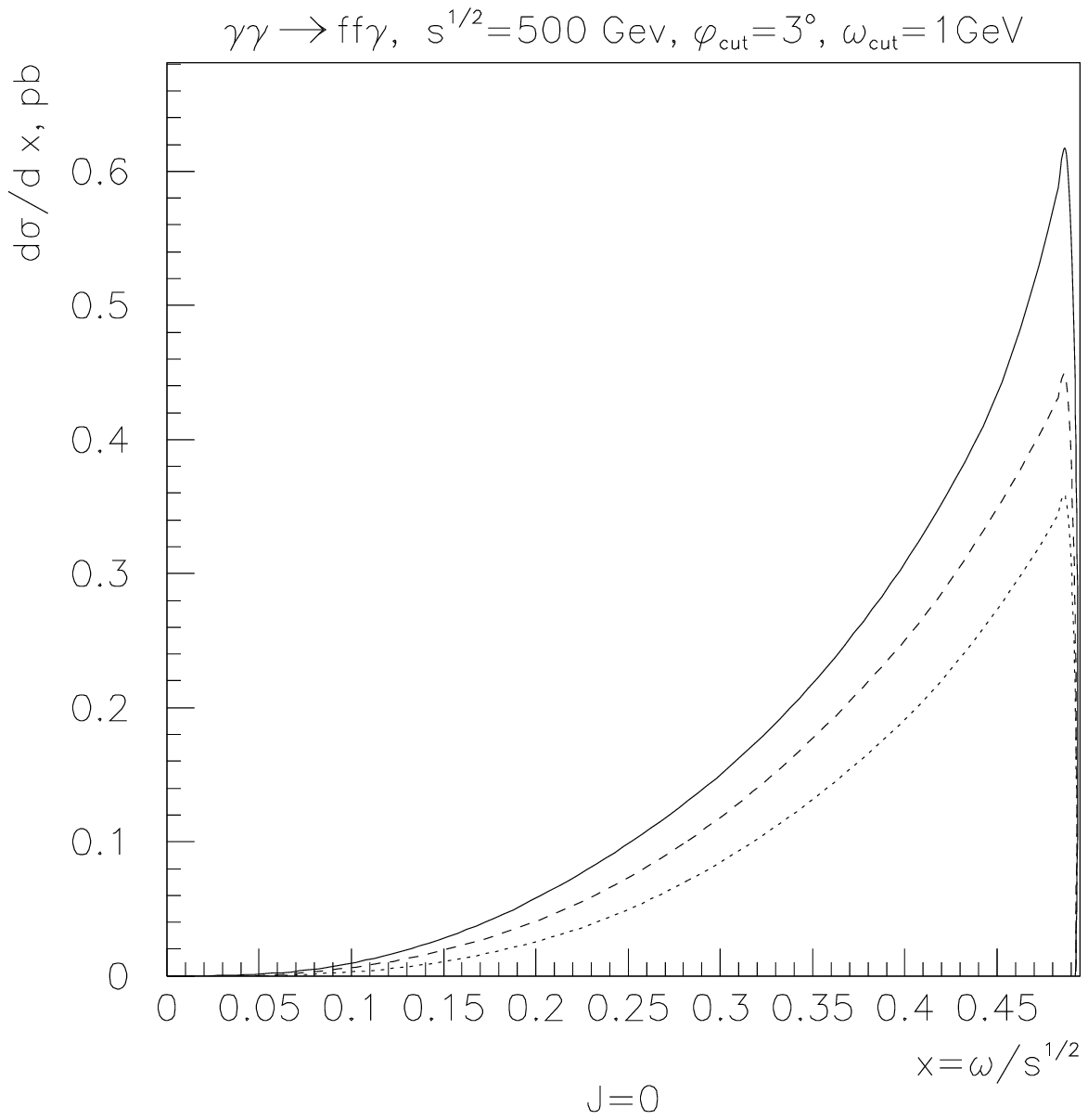}
\label{d_1_500}
\end{minipage}\hfill
\begin{minipage}[b]{.475\linewidth}
\centering
\includegraphics[width=\linewidth, height=3.8in, angle=0]{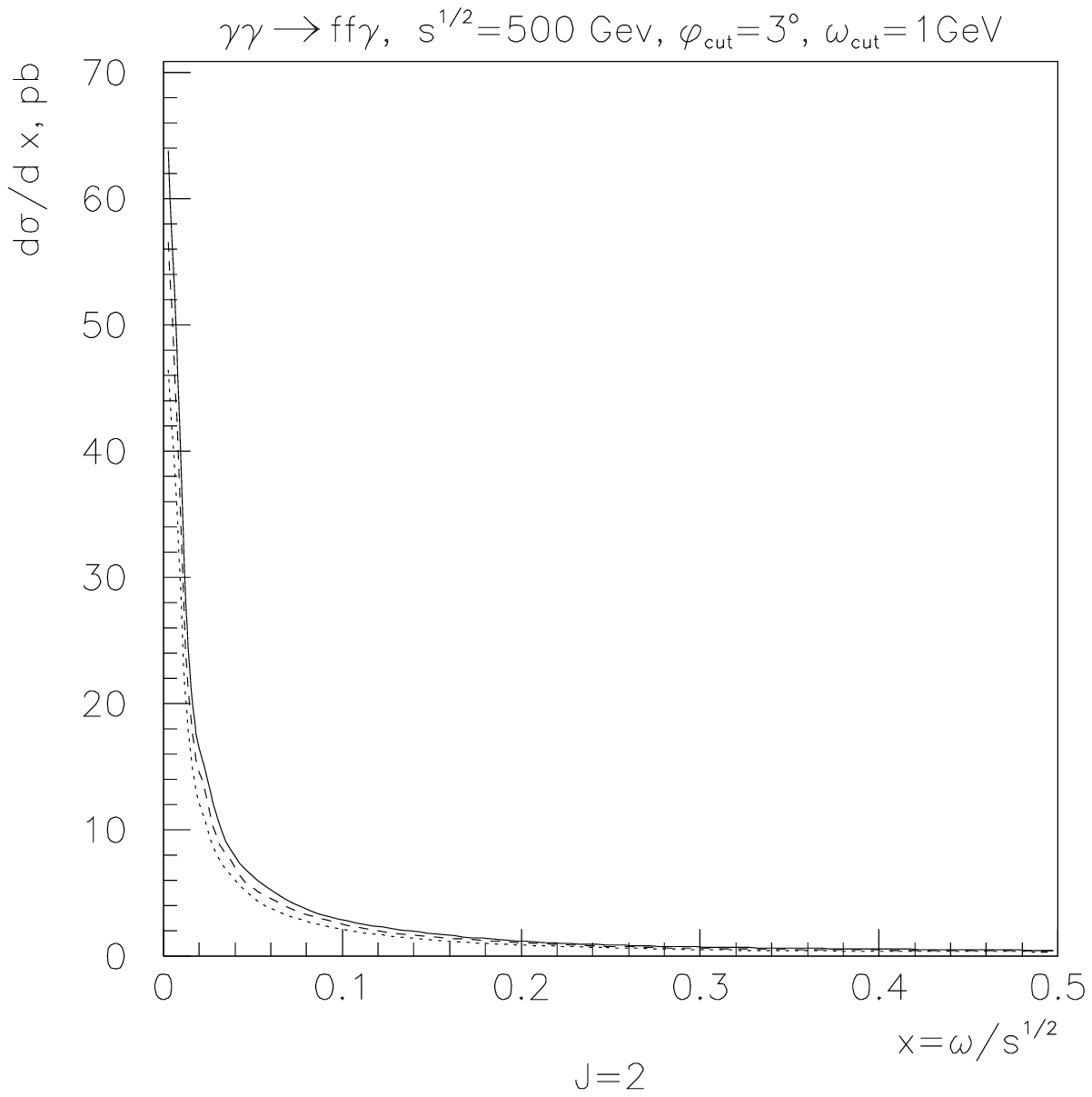}
\label{d_2_500}
\end{minipage}
\caption{The cross section ${d\sigma}\slash{d x}$ for $\sqrt{s} = 500 GeV$.
The cuts are: $5^o,7^o$ and $10^o$ on the polar angle,
$3^o$ on the collinear angle and $1 GeV$ on
the minimal final-state particle energy.
}\label{f2}
\end{figure}

\begin{figure}[h!]
\leavevmode
\begin{minipage}[b]{.33\linewidth}
\centering
\includegraphics[width=\linewidth, height=4.0in, angle=0]{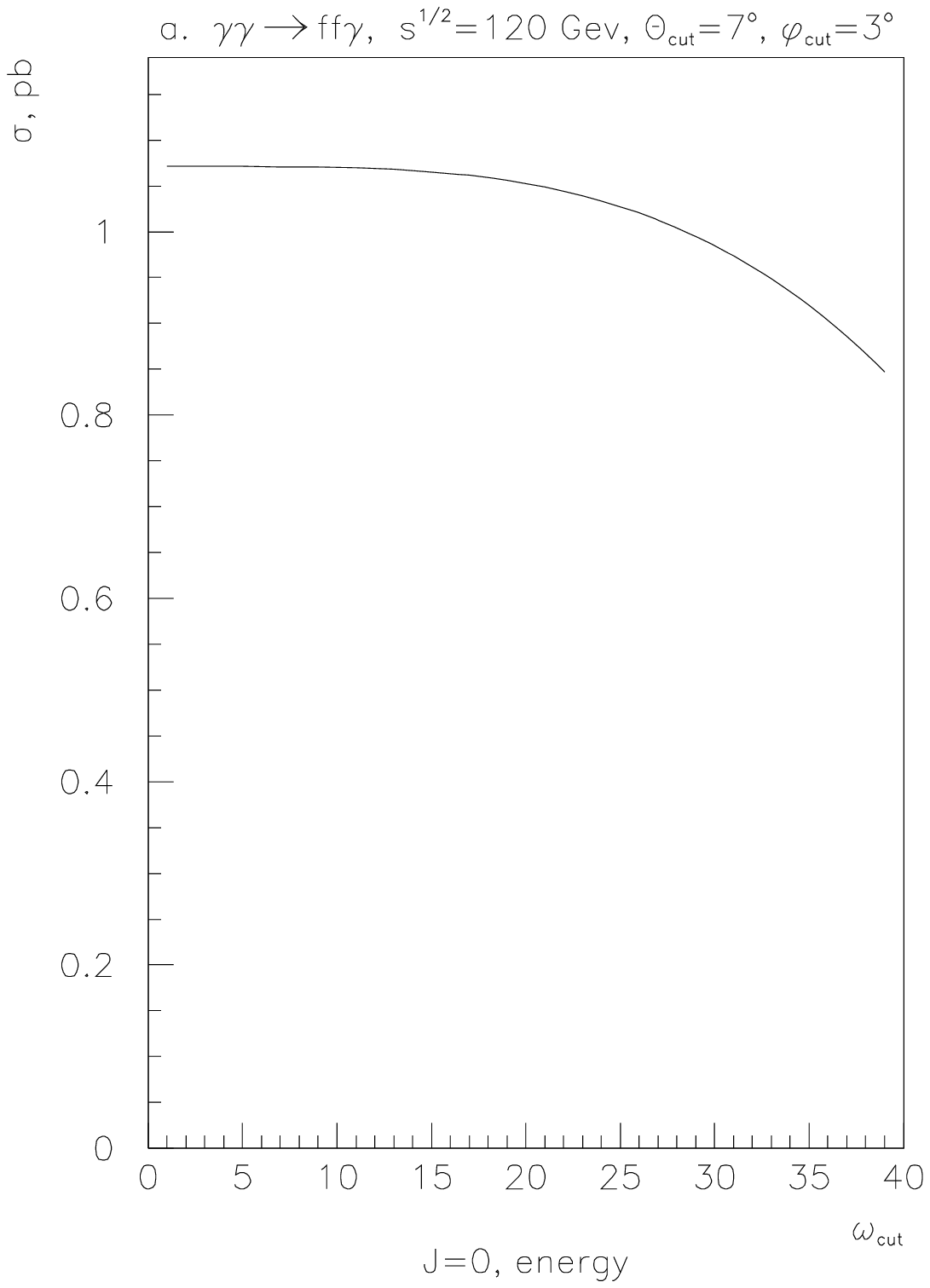}
\end{minipage}\hfill
\begin{minipage}[b]{.33\linewidth}
\centering
\includegraphics[width=\linewidth, height=4.0in, angle=0]{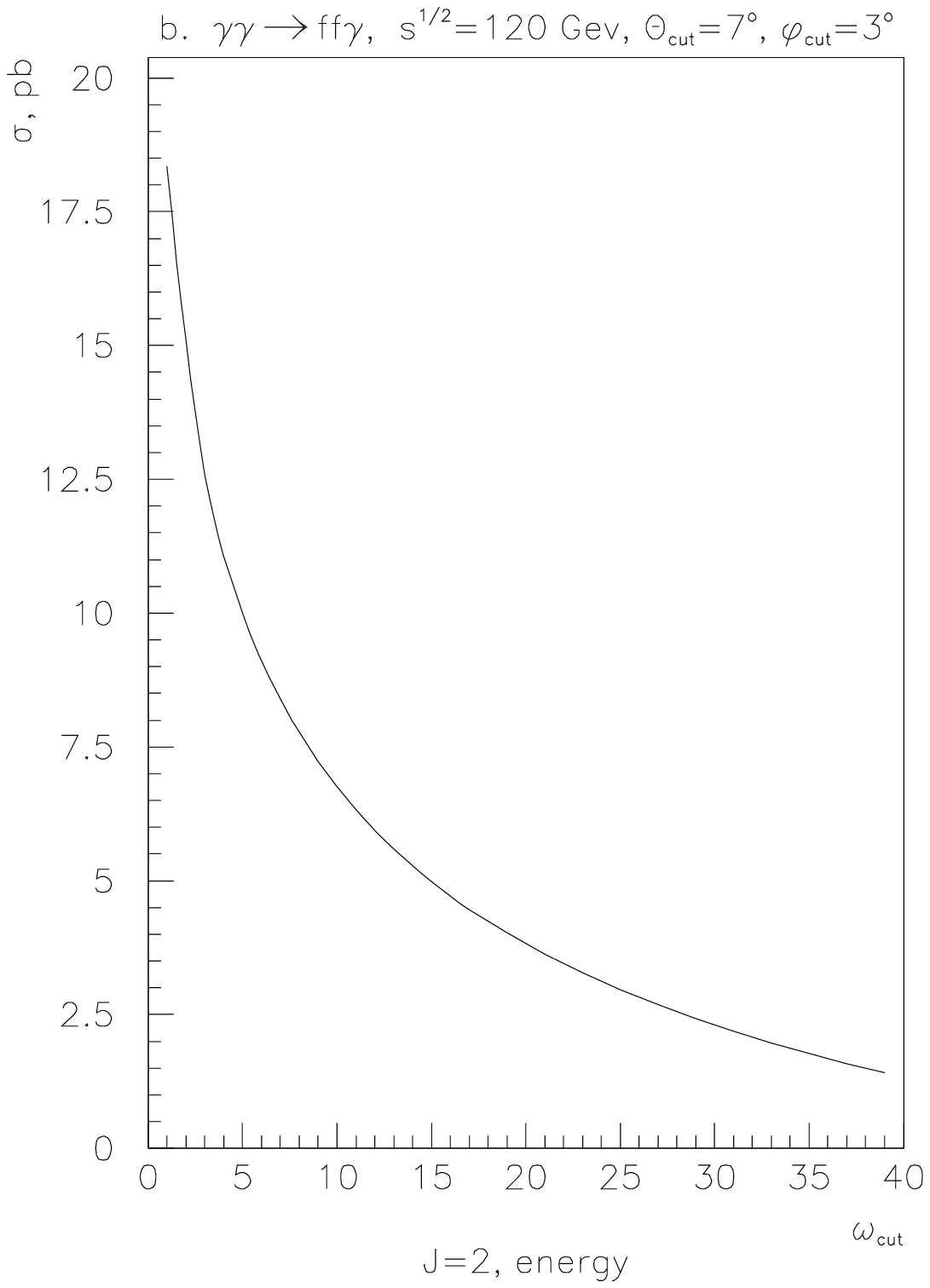}
\end{minipage}
\hfill
\begin{minipage}[b]{.33\linewidth}
\centering
\includegraphics[width=\linewidth, height=4.0in, angle=0]{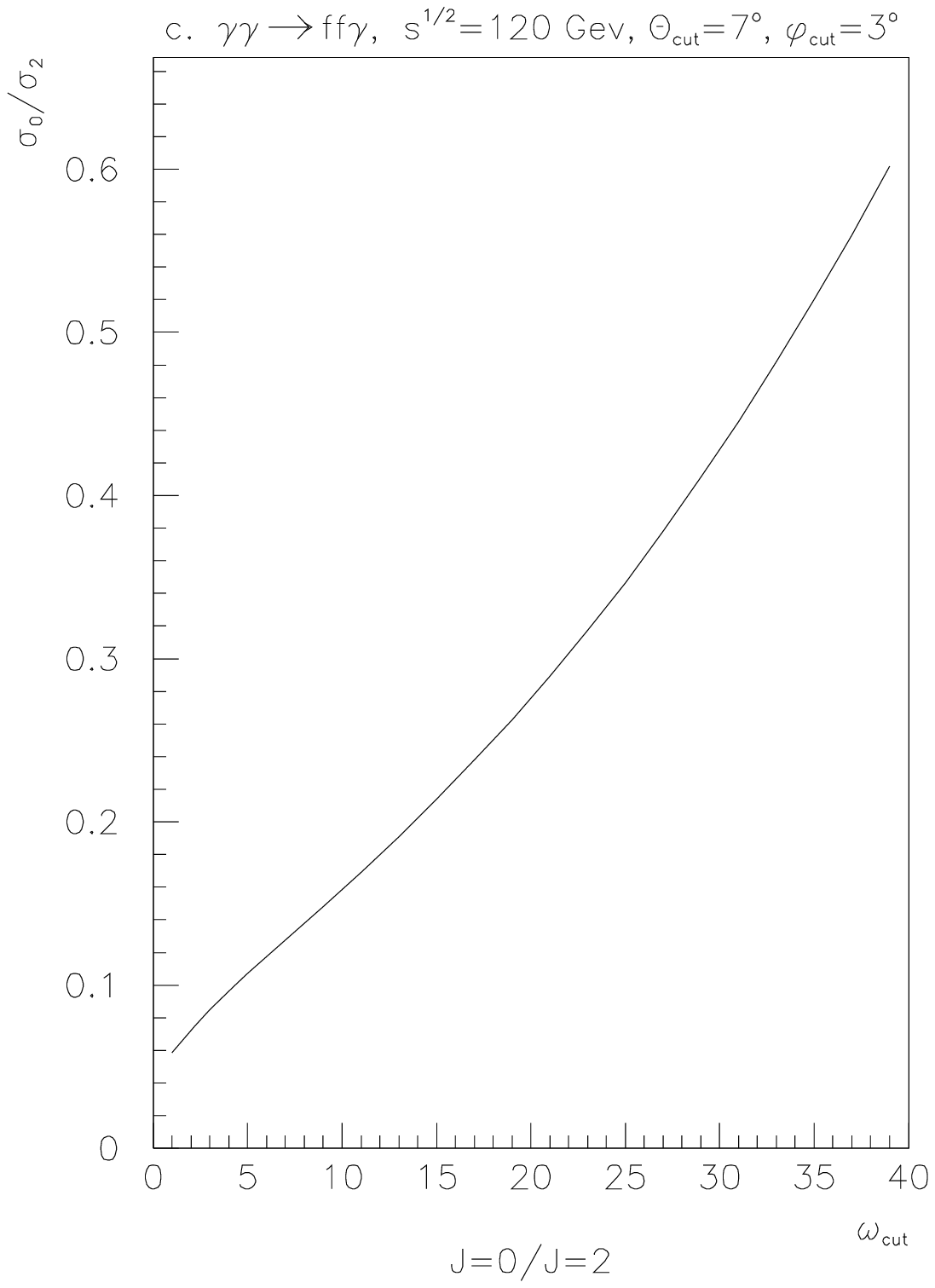}
\end{minipage}
\caption{ Total cross section dependence on the final photon
minimal energy. The cuts are: $7^o$ on the polar angle, $3^o$
between final particles and $1 GeV$ on the minimal fermion energy.
}\label{f31}
\end{figure}

First of all we discuss the energy spectrum of final photons.
As it is presented in fig. \ref{f1}-\ref{f2},
the differential cross section
(${d \sigma}\slash{d x}$)
on J=2 beams
decreases while one on J=0 beams raises with increasing of the
final-state photon energy.
This leads to the conclusion that if one raise the threshold
on $w$,
the process on J=2 beams will be greatly restricted,
but the rate of J=0 events remains almost unchanged.
In fig. \ref{f31} we present the
total cross section dependence on the $w$-cut.
According to fig. \ref{f31}c, one can achive
the ratio $\sigma_{J=0} \slash \sigma_{J=2}$ up to $0.5$
without sufficient decreasing of $\sigma_{J=2}$.

\begin{figure}[h!]
\leavevmode
\begin{minipage}[b]{.33\linewidth}
\centering
\includegraphics[width=\linewidth, height=4.0in, angle=0]{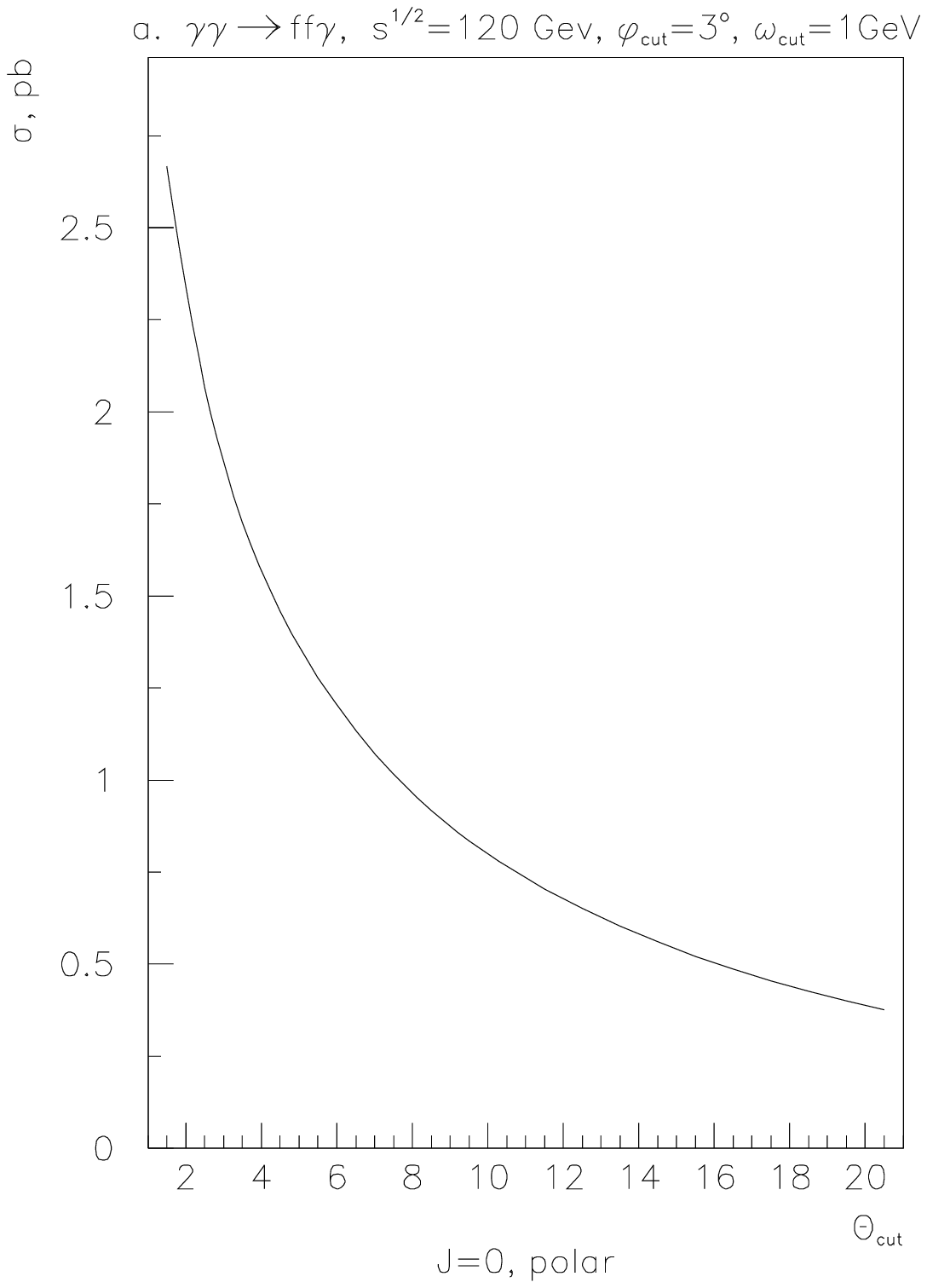}
\end{minipage}\hfill
\begin{minipage}[b]{.33\linewidth}
\centering
\includegraphics[width=\linewidth, height=4.0in, angle=0]{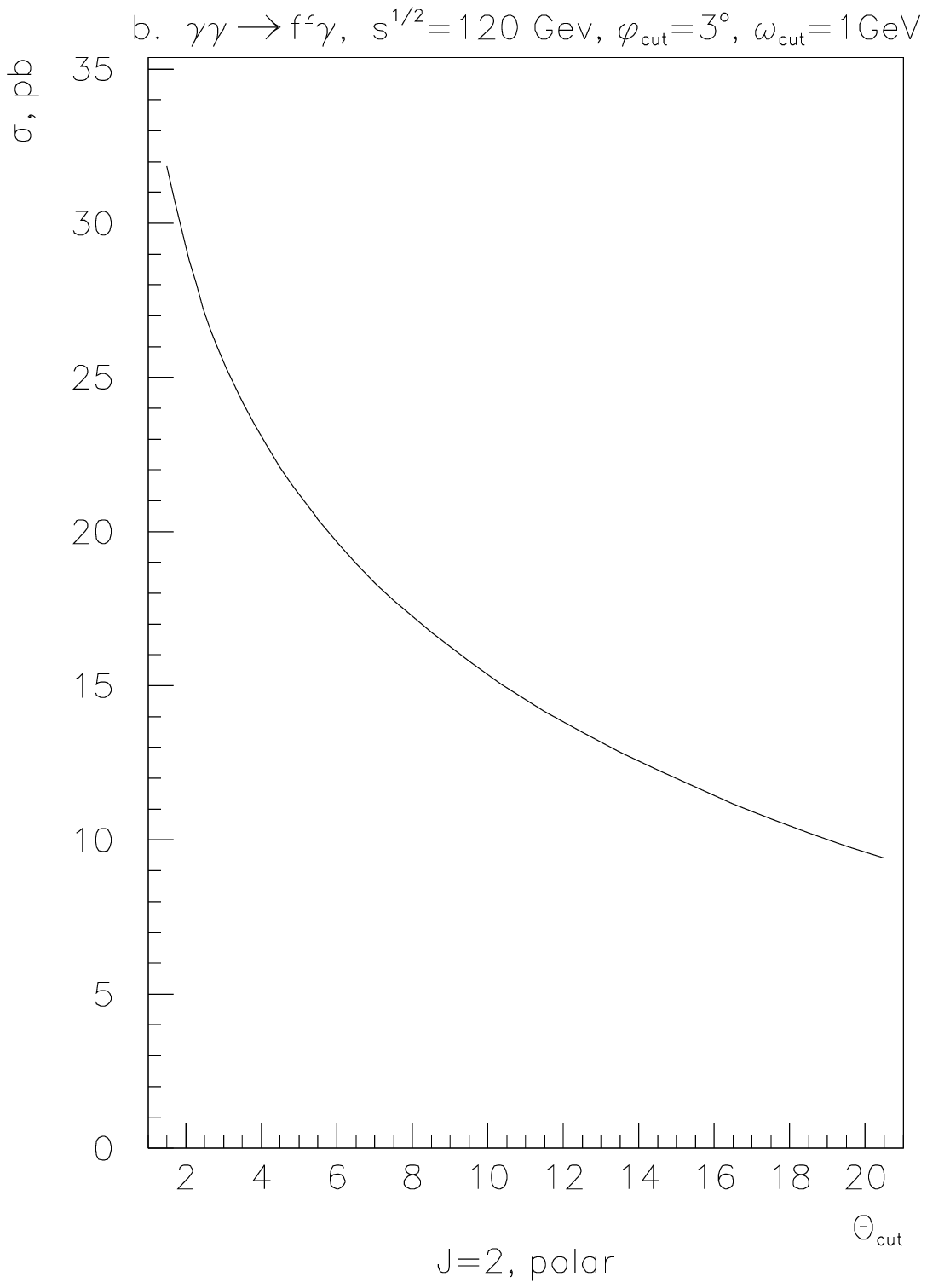}
\end{minipage}\hfill
\begin{minipage}[b]{.33\linewidth}
\centering
\includegraphics[width=\linewidth, height=4.0in, angle=0]{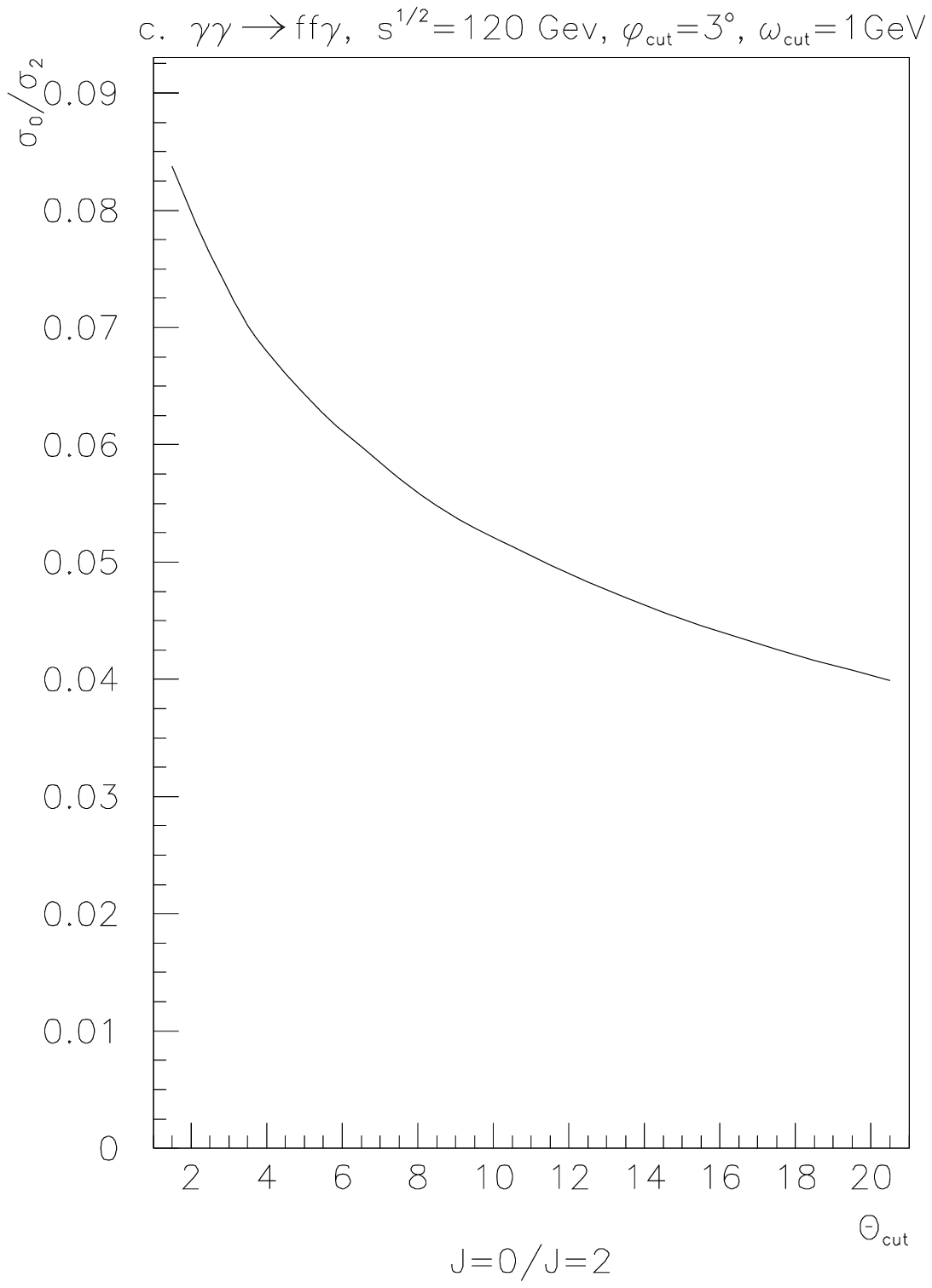}
\end{minipage}
\caption{
Total cross section dependence on the polar angle cut.
The other cuts are: $3^o$ between final particles
and $1 GeV$ on the minimal final-state particle energy.
}\label{f32}
\end{figure}

\begin{figure}[h!]
\leavevmode
\begin{minipage}[b]{.33\linewidth}
\centering
\includegraphics[width=\linewidth, height=4.0in, angle=0]{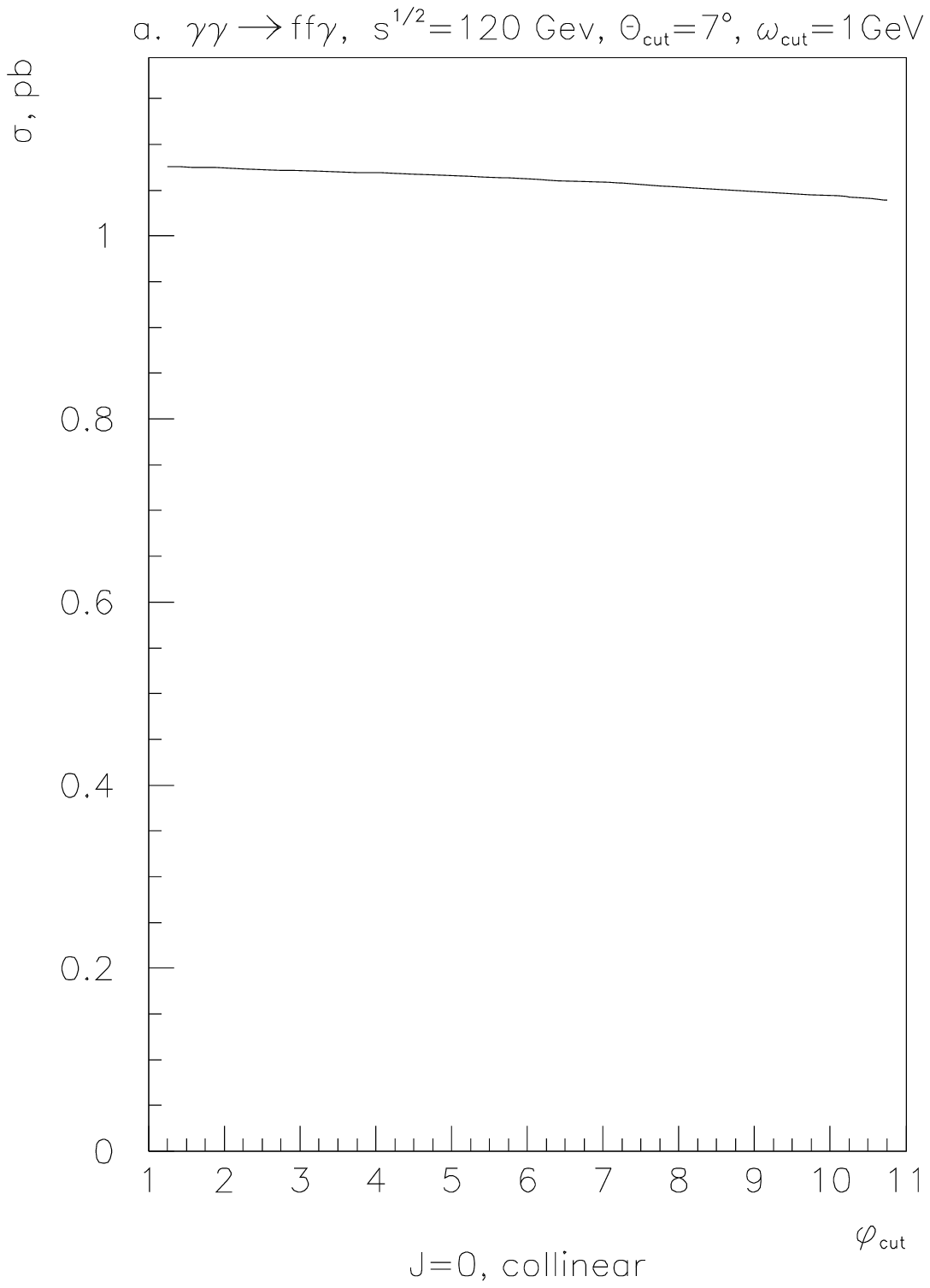}
\end{minipage}\hfill
\begin{minipage}[b]{.33\linewidth}
\centering
\includegraphics[width=\linewidth, height=4.0in, angle=0]{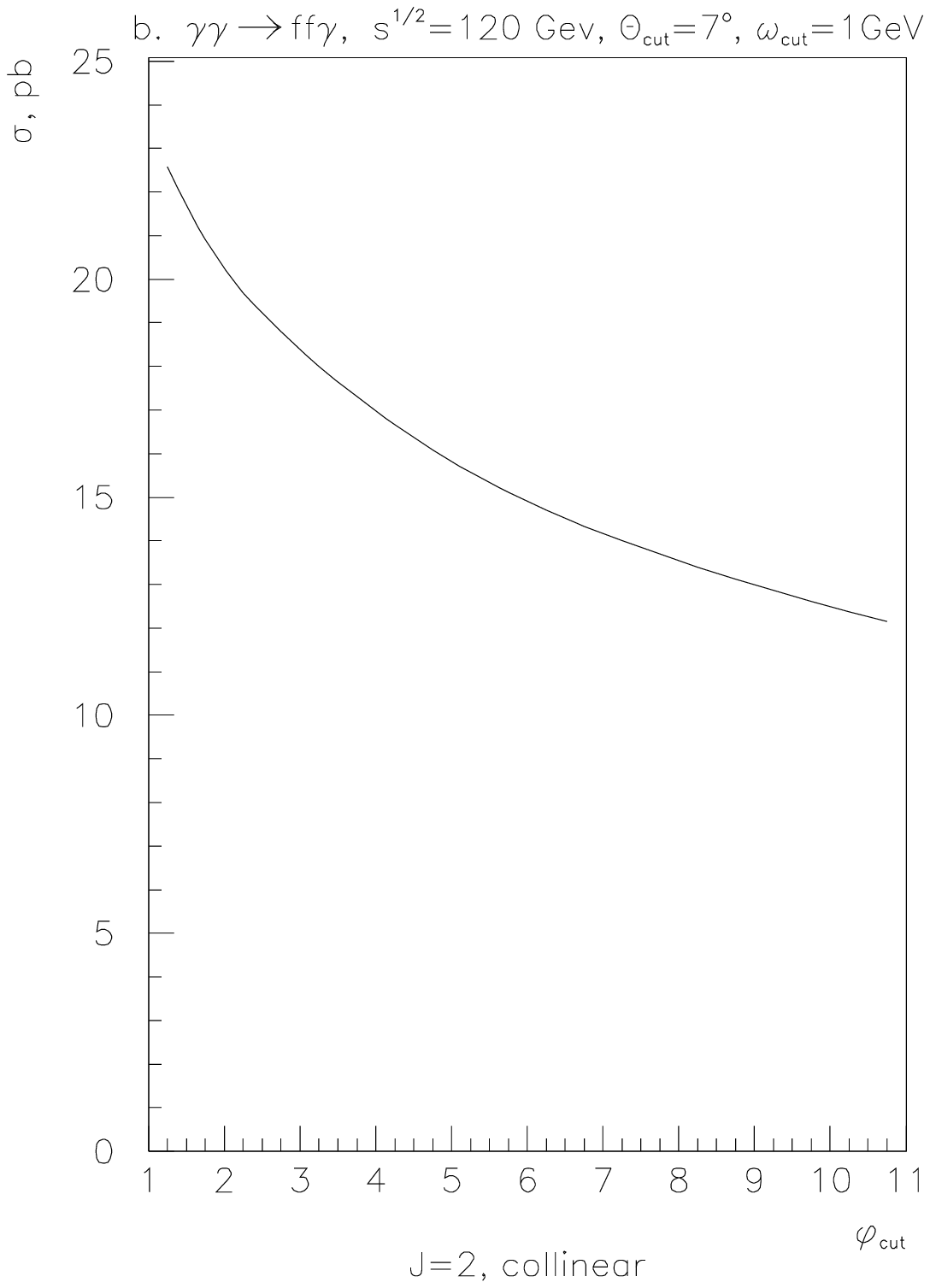}
\end{minipage}\hfill
\begin{minipage}[b]{.33\linewidth}
\centering
\includegraphics[width=\linewidth, height=4.0in, angle=0]{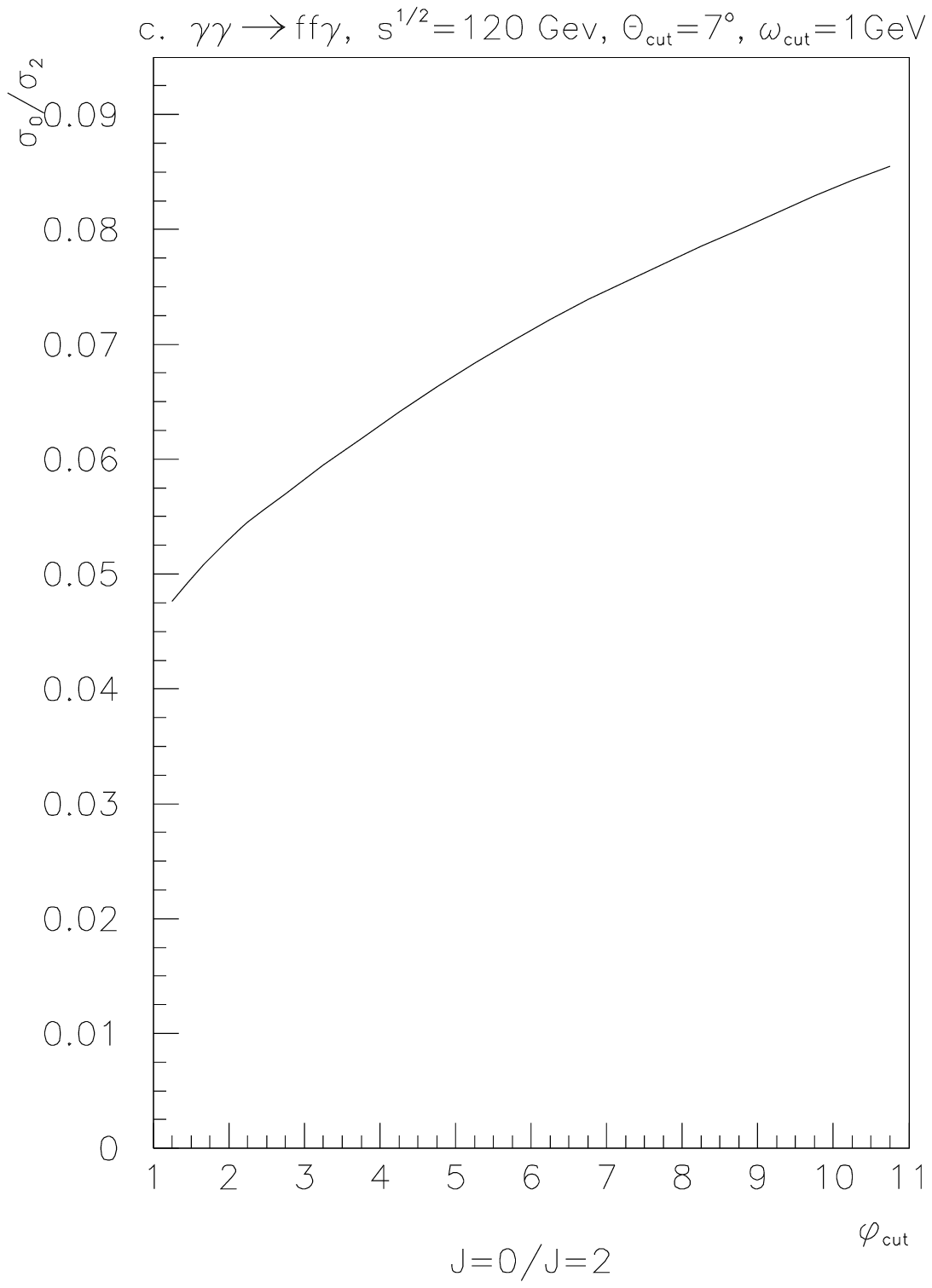}
\end{minipage}
\caption{
Total cross section dependence on the minimal angle between final particles.
The other cuts are: $7^o$ on the polar angle
and $1 GeV$ on the minimal final-state particle energy.
}\label{f33}
\end{figure}

Next we analyse the total cross section dependence on
the angle cuts ($\Theta_{min}$ and $\varphi_{min}$,
see figs. \ref{f32}-\ref{f33}).
In $\gamma^+\gamma^+$- experiments ($J=0$) most of
final fermions are radiated closely to the axis of initial photon
(polar axis) and can't be detected.
The $\gamma^+\gamma^-$- experiments
($J=2$) have the large fraction of particles emitted at large
angles. But they are likely emitted as the collinear
fermion-photon pairs, that also can't be separated in the
detector.
In the figs. \ref{f32}c and \ref{f33}c we
present the ratio of the total cross sections
$\sigma_{J=0} \slash \sigma_{J=2}$
at certain configuration of cuts.
We conclude that it can be increased
by reducing the polar angle cut $\Theta_{min}$,
and by raising the collinear angle cut $\varphi_{min}$.

\begin{figure}[h!]
\leavevmode
\begin{minipage}[b]{.5\linewidth}
\centering
\includegraphics[width=\linewidth, height=4.0in, angle=0]{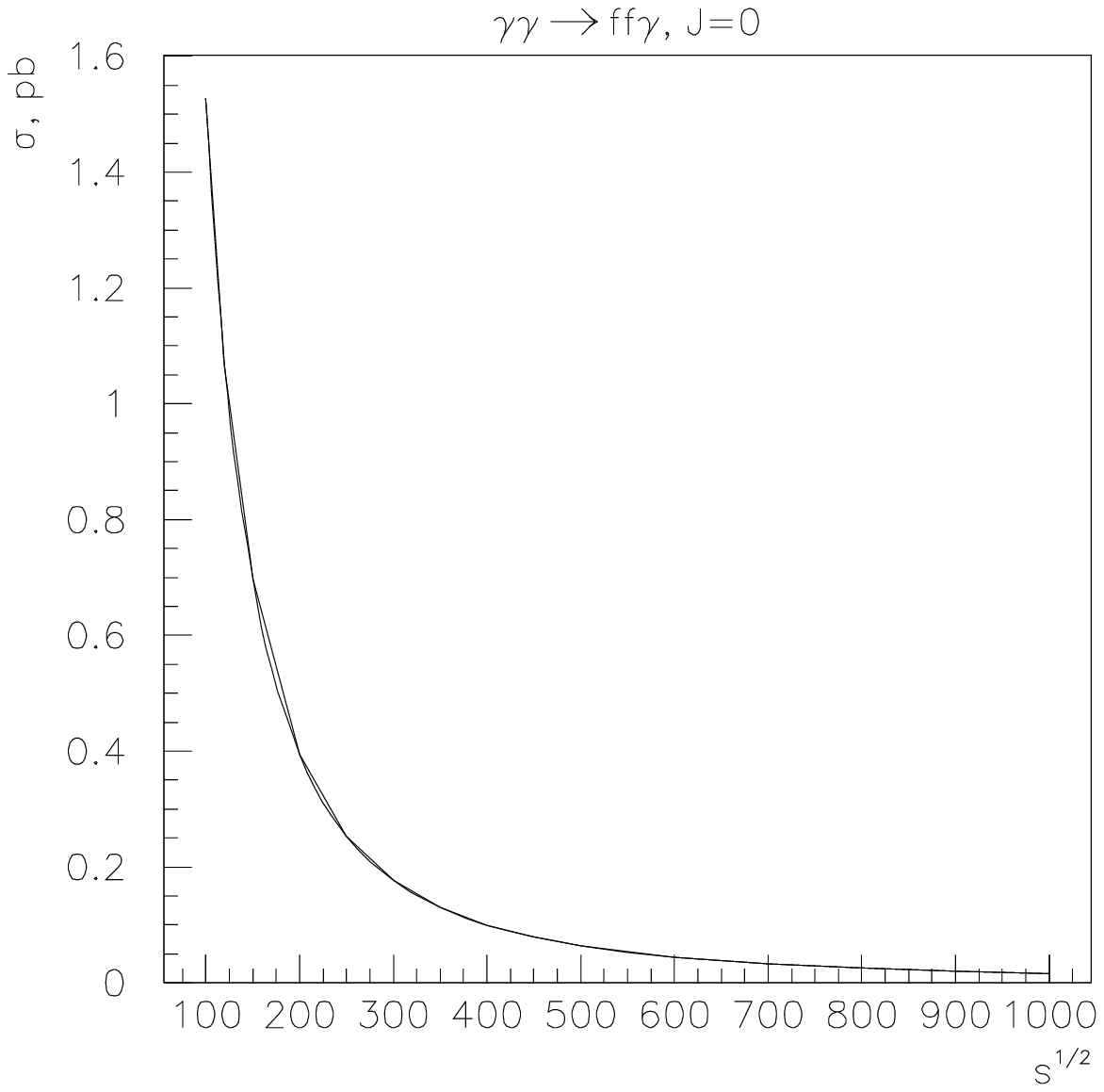}
\end{minipage}\hfill
\begin{minipage}[b]{.5\linewidth}
\centering
\includegraphics[width=\linewidth, height=4.0in, angle=0]{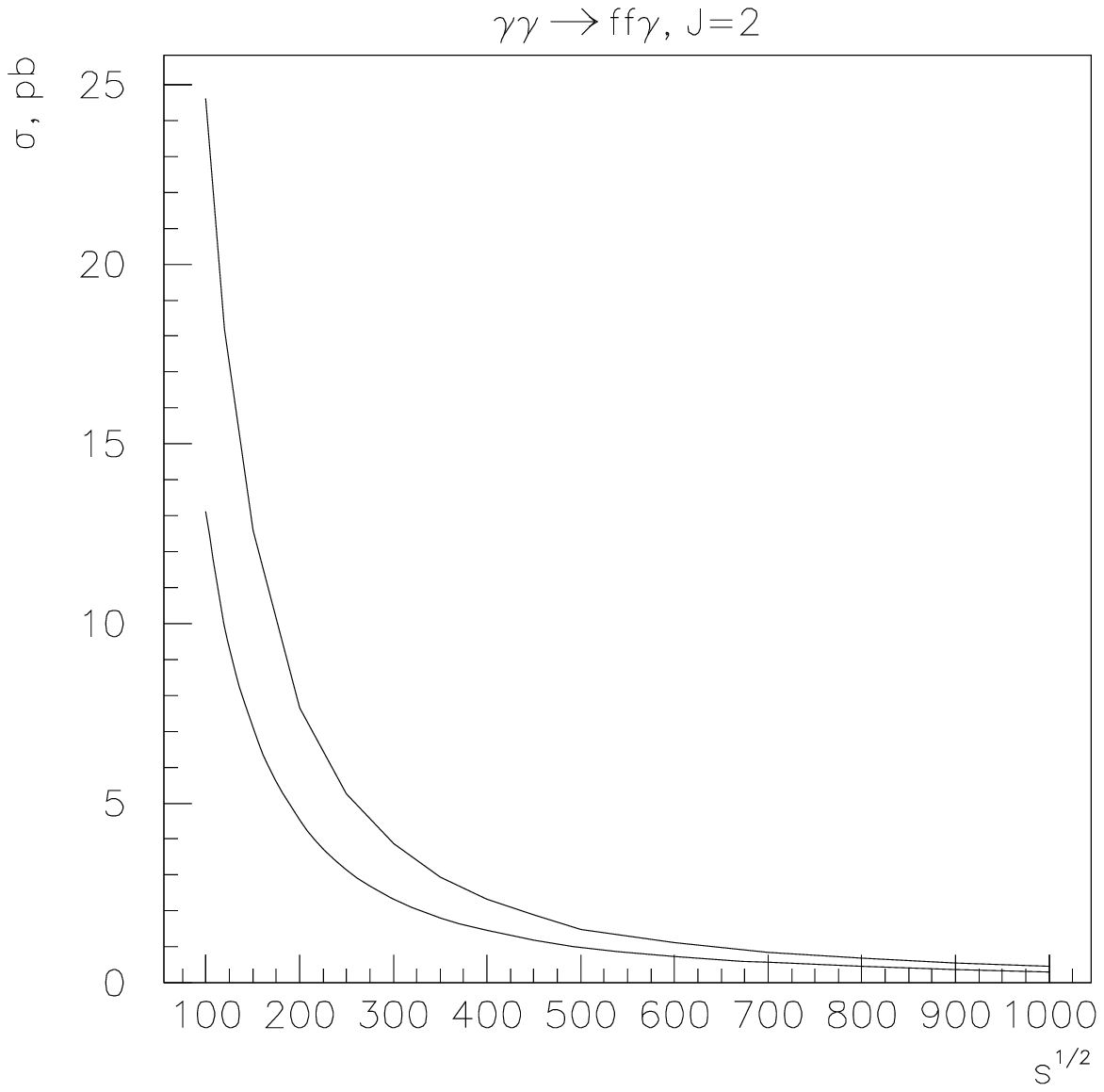}
\end{minipage}
\caption{
The dependence of total cross section on the c.m.s. energy.
The cuts are: $7^o$ (polar angle), $3^o$ (collinear angle),
$1 GeV$ (minimal fermion energy),
$1 GeV$ ($w_{cut}$, the higher line) and $5 GeV$ ($w_{cut}$, the lower line).
}\label{f4}
\end{figure}

In fig. \ref{f4} we demonstrate the total cross section dependence
on the c.m.s. energy.
It is close to the law $1 \slash s$
due to all the invariants in calculations are proportional to $s$.
The only non-linear effect is the $w$-cut.

We discover that
the ratio of events on J=0 and J=2 beams
strongly depends on the experimental cuts.
We obtained the region (the configuration of cuts)
where the processes on the both J=0 and J=2 beams
have the close cross sections.
That is the region of small polar angle cut,
high collinear angle cut
and high minimal energy of final-state photons.
At these parameters the total cross sections
of $ \gamma\gamma\to f \bar{f}\gamma$ in
experiments using $\gamma^+\gamma^+$- and $\gamma^+\gamma^-$- beams
appear to be the same order of magnitude.

\newpage




\end{document}